\documentstyle[aps,prb]{revtex}

\begin{document}

\draft

\title
{Effective mass of one $^{\bf 4}$He atom in liquid $^{\bf 3}$He}
\author{F. Arias de Saavedra}
\address{
Departamento de F\'\i sica Moderna,
Universidad de Granada, E-18071 Granada, Spain}
\author{J. Boronat}
\address{
Departament de F\'\i sica i Enginyeria Nuclear, Campus Nord B4-B5,
\protect\\ Universitat Polit\`ecnica de Catalunya, E-08028 Barcelona,
Spain} \author{ A. Polls}
\address{
Departament d'Estructura i Constituents de la Mat\`eria, Universitat de
Barcelona, \protect\\
Diagonal 647, E-08028 Barcelona, Spain}
\author{A. Fabrocini}
\address{
Department of Physics, University of Pisa,
INFN Sezione di Pisa, I-56100 Pisa, Italy}
%\date{\today}

\maketitle

\begin{abstract}
A microscopic calculation of the effective mass of one $^4$He impurity
in homogeneous liquid $^3$He at zero temperature is performed for an
extended Jastrow--Slater wave function, including two-- and three--body
dynamical correlations and also backflow correlations between
the $^4$He atom and the particles in the medium. The effective mass at
saturation density, $m_4^*/m_4=1.21$, is in very good agreement with the
recent experimental determination by Edwards {\it et al}. The three--
particle correlations appear to give a small contribution to the effective
mass and different approximations for the three--particle
distribution function give almost identical results for $m_4^*/m_4$.
\end{abstract}


%\twocolumn

\narrowtext

Recently, Edwards {\it et al}. \cite{ED92} have used the Zharkov--Silin
Fermi
liquid theory of dilute solutions of $^4$He in normal liquid $^3$He to
determine the chemical potential ($\mu_4$) and the effective mass
($m_4^*$) for the limiting case of one $^4$He impurity. The experimental
input data for this analysis were the recent low temperature
measurements of the phase separation by Nakamura {\it et al}.
\cite{NA90}
The same theory was previously applied to older experimental data by
Laheurte and Saam,
\cite{Saam,LA73} and their predictions for $\mu_4$ and $m_4^{*}$ differ
notably from those of Ref.\onlinecite{ED92}. At zero pressure,
Edwards {\it et al.} report $\mu_4=
-6.95\ K$ and $m_{4}^{*}/m_{4}=1.1$, \cite{ED92} whereas
$\mu_4=-6.60 \ K$ and $m_{4}^{*}/m_{4}=4.5$ in
Refs. \onlinecite{Saam,LA73}.
The disagreement was attributed to the fact that the validity of
the Fermi
liquid theory is ensured at temperatures below $0.1\ K$, \cite{ED92}
while the results of Ref. \onlinecite{Saam,LA73} were obtained
from experimental data at $T \geq 0.5\ K$.
Although the difference between the two experimental values of $\mu_4$
is rather small, our recent microscopic calculations \cite{BF94} seem to
support the more bounded result of Edwards {\it et al}. On the other
hand, the experimental determinations of the impurity effective mass are
appreciably different.
At this point, it is clear that a fully microscopic calculation
of $m_4^*$ would be very enlighting.

In the present work, we evaluate the excitation spectrum and the
effective mass
of a $^4$He impurity in liquid $^3$He using a trial wave function of the
type
\begin{equation}
\Psi_v({\bf k})= \rho_{B}({\bf k}) \, \Psi_0 \, ,
\label{eq:wf}
\end {equation}
where $\Psi_0$ is the ground--state wave function of the $^3$He medium
plus one $^4$He
atom and $\rho_{B}({\bf k})$ is an excitation operator defined as:
\begin{equation}
\rho_{B}({\bf k})= \rho_I({\bf k})~F_{B} \ ,
\label{eq:bf}
\end {equation}
where $\rho_I({\bf k})=\exp (i{\bf {k \cdot r_I}})$ describes the
impurity
travelling through the medium as a plane wave of momentum ${\bf k}$ and
the correlation operator
\begin {equation}
F_{B}=\prod_{i=1}^A f_{B}({\bf k},{\bf r}_{Ii})
\label{eq:bf1}
\end{equation}
incorporates backflow correlations between the impurity $I$ ($^4$He
atom) and the $A$ $^3$He atoms of the bulk.
Backflow correlations have proved to be relevant for a realistic
study of the effective mass of one $^3$He impurity in liquid
$^4$He. \cite{FC56,OW81,FF86} They play also an important role in the
evaluation of the binding energy of pure $^3$He. \cite{MF83,VB88}

The Hamiltonian of the system is written as
\begin {equation}
H(A+1)=H(A)+H_{I}(A+1) \ ,
\label{eq:h1}
\end{equation}
where
\begin {equation}
H(A)=- \frac {\hbar^2}{2 m_3} \sum_{i=1}^A \nabla_{i}^{2} +
\sum_{i<j}^{A} V(r_{ij}) \label{eq:h2}
\end{equation}
is the Hamiltonian of the pure $^3$He background, and
\begin{equation}
H_{I}(A+1)= -\frac {\hbar^2}{2 m_4} \nabla_I^{2}+
\sum_{i=1}^{A}V(r_{iI})
\label{eq:h3}
\end{equation}
are additional terms related to the impurity.

The variational approach starts with the choice of the trial wave
function $\Psi_0$. As in our previous paper, \cite {BF94} we take an
extended Jastrow--Slater wave function for the $A+1$ particles:
\begin{equation}
\Psi_0= F_J \, F_T \, \phi(1,...,A)\ ,
\label{eq:wf1}
\end{equation}
where $\phi(1,...,A)$ is the Fermi gas wave function for the $A$ $^3$He
atoms,
$F_J$ is a Jastrow correlation operator embodying two--body dynamical
correlations
\begin{equation}
F_J=\prod_{i<j}^{A} f^{(3,3)}(r_{ij}) \prod_{i=1}^{A} f^{(3,I)}
(r_{iI}) \ ,
\label{eq:wf2}
\end{equation}
and the triplet correlation operator $F_{T}$ is written as:
\begin{equation}
F_{T}=\prod_{i<j<k}^{A} e^{-q_{ijk}/2} \prod_{i<j}^{A}
e^{-q_{Iij}/2} \ ,
\label{eq:tr1}
\end{equation}
with
\begin{equation}
q_{\alpha j k}=\sum_{cyc}\xi(r_{\alpha j}) \xi(r_{\alpha k}) {\bf
r}_{\alpha j} \cdot {\bf r}_{\alpha k}\ .
\label{eq:tr2}
\end{equation}
Here, $\sum_{cyc}$ denotes a summation on the cyclic permutations of the
indices $\alpha,j,k$, where the index $\alpha$ may either represent a
$^3$He atom or the $^4$He impurity. The triplet correlations have been
found to be very important to properly describe the equation of state of
the pure phase \cite{MF83,VB88,Kal} and to calculate the chemical
potential of the $^4$He impurity. \cite{BF94}

The expectation value of the Hamiltonian with respect to $\Psi_{v}({\bf
k})$ is given by
\begin{equation}
E^{v}(k)= E_{0}^{v}+ \frac {\left \langle \Psi_{0} \left |
\rho_B^{\dagger} \, [H,\rho_{B}]
\right |
\Psi_0\right \rangle}
{\left\langle \Psi_0 \mid \Psi_{0} \right \rangle}
\label{eq:1}
\end {equation}
where we have taken advantage of the unitary character of the excitation
operator $\rho_{B}({\bf k})$.
If one ignores backflow correlations between the impurity and the
medium, by assuming  $\rho_{B}({\bf k})=\rho_{I}({\bf k})$,
a simple parabolic spectrum for $E^{v}(k)$ is obtained:
\begin{equation}
E^{v}(k)=E_{0}^{v}+ \frac{\hbar^2 k^2}{2 m_4} \ ,
\label{eq:spec1}
\end{equation}
where $E_0^v=\langle \Psi_0 | H | \Psi_0 \rangle$. In this case, the
$^4$He  effective mass is equal to the bare mass,
pointing to an excessively simple choice for the excitation operator.

A better ansatz is given by $\Psi_v({\bf k})$ of Eq. (\ref{eq:wf}) which
explicitly contains backflow correlations between the impurity and the
$^3$He atoms. In particular, the backflow correlation operator has been
taken of the form \cite{OW81,FF86}
\begin{equation}
F_{B}=\prod_{i=1}^{A} \exp \left[ i{\bf k \cdot r}_{Ii} \,
\eta(r_{Ii})\right] \ .
\label{eq:bf2}
\end{equation}

As the interatomic potential depends only on the relative distance
between
the atoms, it commutes with $\rho_B({\bf k})$. Therefore, in
Eq.(\ref{eq:1}) it is only necessary
to consider the commutator with the kinetic energy operator.
After some integration by parts,
a generic contribution of the kinetic energy operator to Eq.(\ref{eq:1}) may be
expressed through the following relation:
\begin{eqnarray}
X_\alpha  =  {\frac {\left \langle \Psi_{0} \left | \rho_B^{\dagger} \,
[{\nabla}^2_\alpha,\rho_{B}]
\right | \Psi_0\right \rangle} {\left \langle \Psi_0 \mid \Psi_{0}
\right \rangle}  }= X_\alpha^\rho+ X_\alpha^\phi \ ,
\label{eq:2}
\end {eqnarray}
with
\begin{eqnarray}
X_\alpha^\rho  = - {\frac {\left \langle \Psi_0  \left | \left (
\roarrow{\nabla}_\alpha
\rho_{B}^{\dagger} \right ) \left ( \roarrow{\nabla}_\alpha
\rho_{B}
\right ) \right | \Psi_{0} \right \rangle} { \left \langle \Psi_0 \mid
\Psi_0 \right \rangle } } \ ,
\label{eq:2a}
\end{eqnarray}
and
\begin{eqnarray}
X_\alpha^\phi  & = &  \frac{\left \langle \phi \left |(F_J F_T)^2
\rho_{B}^{\dagger} \left ( \roarrow{\nabla}_\alpha \rho_{B} \right )
\roarrow{\nabla}_\alpha \right | \phi \right \rangle}
{\left \langle \Psi_0 \mid \Psi_0 \right \rangle }  \nonumber \\
 &  & -\frac{\left \langle \phi \left |\loarrow{\nabla}_\alpha \,
\rho_{B}^{\dagger}\left ( \roarrow{\nabla}_\alpha \rho_{B} \right )
(F_J F_T)^2\right |\phi \right \rangle}
{\left \langle \Psi_0 \mid \Psi_0 \right \rangle}  \ ,
\label{eq:2b}
\end{eqnarray}
where the subscript $\alpha$ labels the generic particle.
The arrows indicate
in which direction ( right or left)  the derivatives are acting.

The first term $X_\alpha^\rho$ (\ref {eq:2a}) is analogous to the
expression
obtained in the case of the $^3$He impurity in liquid $^4$He.
\cite{OW81,FF86} The second one $X_\alpha^\phi$ (\ref{eq:2b})
directly originates
from the Fermi character of the $^3$He medium, as it comes from
the kinetic energy operator acting on the Slater determinant
$\phi$. Clearly, $\alpha=I$ does not contribute to $X_\alpha^\phi$.
Moreover, by inspecting the cluster expansion of $X_\alpha^\phi$,
it results to be strictly zero. In fact, for direct cluster terms,
where the
$\alpha$--particle is not exchanged, each of the two pieces of
$X_\alpha^\phi$ is zero, after summing over the momentum carried by
$\alpha$. Terms in which $\alpha$ is exchanged cancel because each of them
gives the same contribution in both pieces.

The explicit expression for the impurity single--particle excitation
energy, measured with respect to the $^4$He chemical potential, is then
\begin{eqnarray}
 \varepsilon_k=E^{v}(k)-E_{0}^{v}= \frac{\hbar^2 k^2}{2 m_4}
 \left [ 1+ e_{2}+ \frac{m_4}{\mu}e_{m}+e_{3}\right],
\label{complexe}
\end{eqnarray}
where
\begin{equation}
e_{2} =  \rho \int d
{\bf r}_{Ij} \, g^{(2)}_{Ij} \left( 2 \eta_{Ij} + \frac{2}{3} r_{Ij}
\eta^{\prime}_{Ij} \right )    \, ,
\label{e2}
\end{equation}
\begin{equation}
e_m = \rho \int d {\bf r}_{Ij} \, g^{(2)}_{Ij}
\left[
\eta_{Ij}^2+\frac{1}{3} \left( r^2_{Ij} (\eta^{\prime}_{Ij})^2 + 2
\eta_{Ij} r_{Ij} \eta^{\prime}_{Ij} \right) \right]  \, ,
\label{em}
\end{equation}
and
\begin{eqnarray}
e_3 & = & \rho^2 \int d {\bf r}_{Ij} d {\bf r}_{Ik} \, g^{(3)}_{Ijk}
\bigg[ \eta_{Ij} \eta_{Ik}    \nonumber \\
 & & \left. + \frac{1}{3} \left( r_{Ij} \eta^{\prime}_{Ij}
\eta^{\prime}_{Ik} r_{Ik} ( \hat{{\bf r}}_{Ij} \cdot \hat{{\bf r}}_{Ik}
)^2 + 2 \eta_{Ij} \eta^{\prime}_{Ik} r_{Ik} \right) \right]\ .
\label{three}
\end{eqnarray}
$\mu$ is the reduced mass ($\mu^{-1}=m_3^{-1}+m_4^{-1}$) and
$g^{(2)}_{Ij}$ and $g^{(3)}_{Ijk}$ are the two-- and three--body
distribution
functions between the impurity and the $^3$He atoms of the medium.
They are the only quantities
carrying information about the antisymmetry of the $^3$He bulk.
It is worthwile to remind that,
by changing $m_4$ with $m_3$ in Eq. (\ref{complexe}), one recovers the
expression for the reverse problem of one $^3$He impurity in liquid
$^4$He (Eq. 2.26 of Ref. \onlinecite{FF86} ), with the obvious
substitution of the appropriate distribution functions.

All the calculations presented in this paper have been performed in the
framework of the so called Average Correlation Approximation (ACA). In
this approximation one considers the same dynamical correlation
functions for all the pairs and triplets in the system, not distinguishing
between the two isotopes. This assumption relies on
the fact that the interatomic potential is the same for all the pairs.
 The drawbacks of the ACA in the evaluation of the
chemical potential of the $^4$He impurity have been extensively
discussed in Ref. \onlinecite{BF94}.

We have used the interatomic Aziz potential \cite{AZ79}
and the two--body correlation factor $f(r)$ has been taken of the
McMillan type:
\begin{equation}
f(r)=\exp \left[ - \frac{1}{2} \, \left( \frac{b}{r} \right )^5 \right]
\ .
\label{mcmill}
\end{equation}
The variational parameter $b$ has been fixed by means of a numerical
minimization of the energy of pure liquid $^3$He. The value
$b=1.15
\ \sigma$ $(\sigma=2.556$ \AA), determined at the $^3$He experimental
equilibrium density
($\rho_0^{exp}=0.277 \ \sigma^{-3}$), has been used for all the
densities. The function $\xi(r)$ of the triplet correlation
(\ref{eq:tr2}) has the same parametrized form used in pure phase
calculations \cite{MF83,VB88,Kal}:
\begin{equation}
\xi(r)= \sqrt{\lambda_t} \exp \left[ - \left( \frac{r-r_t}{\omega_t}
\right )^2 \right ] \ .
\label{xi}
\end{equation}
The density dependence of the triplet variational parameters is
neglected and the optimum values at $\rho_0^{exp}$ have been used
everywhere. These values are $\lambda_t=-0.75 \ \sigma^{-2}$, $r_t=0.85\
\sigma$ and $\omega_t=0.45\ \sigma$. \cite{ARIAS}

The distribution functions have been computed by using the Fermi HyperNetted
Chain (FHNC) technique, in the so called FHNC/S(T) approximation
\cite {MF83}
to take into account the elementary diagram ( and triplet) contributions.
As reported in Ref. \onlinecite{BF94}, the chemical potential provided
by the
variational wave function $\Psi_0$, at the FHNC/ST saturation density
($\rho_0=0.252 \ \sigma^{-3}$), is $-6.60\ K$.

The function $\eta(r)$ (\ref{eq:bf2}), adopted for the backflow
correlation, is of the form \cite {OW81,FF86,MF83,VB88,ARIAS}
\begin{equation}
\eta(r)=A_0 \exp \left[ - \left( \frac{r-r_0}{\omega_0} \right) ^2
\right ] \ .
\label{eta}
\end{equation}
In our case, the backflow parameters $A_0=0.2$, $r_0=0.8 \ \sigma$
and $\omega_0=0.375 \ \sigma$, taken from Ref. \onlinecite{ARIAS},
are used at all the densities.

As the single--particle spectrum (\ref{complexe}) is quadratic in $k$,
the effective mass is given by
\begin {equation}
\left(\frac {m_{4}^{*}}{m_4}\right)^{[\beta]} =
 \frac {1} {1+ e_2 + (m_4/\mu)\, e_m + e_{3}^{\beta} } \ ,
\label{mass}
\end{equation}
where $\beta$ labels the
approximation used in evaluating the three--body distribution function.

Table I reports the effective mass  obtained at the experimental
equilibrium density $\rho_0^{exp}$ in different approximations. Also
given is the value of the two--body contribution $(m_4^*/m_4)^{[2]}$,
i.e. taking $e_3^\beta=0$ in Eq. (\ref{mass}).

The Jastrow (FHNC/S) and the Jastrow plus
Triplet correlation (FHNC/ST) models give nearly the same results. The
three-body
distribution function has been evaluated in the Kirkwood superposition
approximation (KSA), in the convolution approximation (CA) and including
the Abe terms (KSA+ABE). \cite {MF83} As shown in the
Table, the three different approximations to $g^{(3)}_{Ijk}$ give very close
results.

The effective mass  at $\rho_0^{exp}$ in FHNC/ST,  with the inclusion of
the Abe diagrams, turns out to be 1.21. This result is
in very good agreement with the most recent experimental determination
$m_{4}^{*}/m_{4}=( 1.1 +0.4\, /  -0.1)$. \cite{ED92}
As it has been pointed out by Leggett, \cite{Leg} the effective mass of
one impurity in a Fermi liquid is always larger than the bare mass.
In FHNC/ST, $e_2=-0.37$, $e_m=0.08$ and the three--body term
(\ref{three}),
in all approximations, is very small, $e^{\beta}_3\simeq 0.01$.
Therefore,  the denominator of Eq. (\ref {mass}) is smaller
than unity, providing an effective mass larger than one.

The HNC/S results have been obtained by setting $\phi_0=1$ in the wave
function, i.e. by treating the $^3$He as a bosonic fluid.
 The comparison with the FHNC/S results indicates that, at this
density,
the influence of the Fermi character of the medium  on the
calculation of the effective mass of the impurity is nearly negligible.

The density dependence of the calculated effective mass is reported in
Table II and  it is also shown in Fig. 1 (full triangles). As one
can see, the effective mass increases linearly with density.

It is also
interesting to compare our results for $m_4^*/m_4$ with the
effective mass of one $^3$He impurity in liquid $^4$He. As there are not
exchange or spin correlations between the two isotopes, one expects
the effective mass to be driven mainly by the density. To deeper explore
this hypothesis we have plotted in Fig. 1 the density dependence of the
effective mass of a $^3$He impurity in liquid $^4$He, for both the
experimental data (full circles) \cite{Expm3} and
the theoretical estimates, obtained by using
 backflow correlations (empty circles).
\cite{FF86} The
density dependence is in both cases approximately linear. Although the
slopes are
different, the extrapolated values of  $m_3^*/m_3$, at the $^3$He
saturation
density, are similar and close to the present evaluation of $m_4^*/m_4$.
As it has been mentioned before, by taking the proper
mass factor $m_3/\mu$ in front of $e_m$,
Eq. (\ref{complexe}) is approximately valid for
one $^3$He impurity in liquid $^4$He, since the differences in the
distribution functions in the two cases are small. \cite{BPF92}
In fact, if one performs a calculation at  $\rho_0^{exp}$
using the mass factor $m_3/\mu$, then  the HNC/ST
result (shown by an open diamond in Fig. 1)
coincides with the backflow extrapolated value. The small difference
between this value and the proper result (full triangle) of $m_4^*/m_4$
arises almost completely from the different mass factors
in front of $e_m$.
The difference practically coincides with the estimate
\begin {eqnarray}
\Delta \left(\frac{m_I^{*}}{m_I} \right) & = &
 \frac {1} {1+ e_2 + (m_3/\mu)\, e_m + e_{3} } \nonumber \\
  & & -\frac {1} {1+ e_2 + (m_4/\mu)\, e_m + e_{3} } = 0.073 \ ,
\label{Dmass}
\end{eqnarray}
obtained by considering the same distribution functions in the two
systems. Assuming a linear extrapolation, the experimental value for
$m_3^*/m_3$
(solid line) is close to the backflow extrapolation (long--dashed line)
at $\rho_0^{exp}$,
pointing out that possible perturbative corrections, beyond the backflow
terms, are small at this low density. These corrections have been
evaluated \cite{FF86}
in Correlated Basis Function theory (CBF) for $m_3^*/m_3$ in
$^4$He. They result to be about 0.5 at the $^4$He saturation
density ($\rho=0.365\ \sigma^{-3}$) and rapidly decreasing with the
density.

It is worthwile to notice that the effective mass of a $^3$He atom
at the Fermi
surface of pure $^3$He ($m^*/m=2.8$) \cite{Gre} is much larger than
the effective mass of the
$^4$He impurity. The statistics and the spin effects, which
are suppressed in the case of the $^4$He impurity, appear to be
the main responsibles for this difference.

To briefly summarize, we have calculated the effective mass of one
$^4$He impurity in liquid $^3$He by using backflow correlations. These
correlations
provide for an accurate description of the $^4$He impurity spectrum at
low momenta. Our results
support the new experimental determination of Edwards {\it et al.}
\cite{ED92} and are far from the previous result of Laheurte and Saam.
\cite{Saam,LA73}

This research was supported in part by DGICYT (Spain) Grant
Nos. PB92-0761, PB90-06131, PB90-0873 and the agreement DGICYT
(Spain)--INFN (Italy).

\begin{figure}
\caption{Density dependence of the impurity effective mass. Full
and open circles
are respectively the experimental data and the backflow results for
$m_3^*/m_3$. The open diamond is the backflow result for $m_3^*/m_3$
at $\rho_0^{exp}$, indicated by an arrow. The full triangles are the
backflow results for $m_4^*/m_4$. The experimental result is plotted as
a full diamond with its error bar. The lines are linear fits to the
corresponding points.
}
\label{fig:resul}
\end{figure}

\begin{table}[]
\caption{$^4$He effective mass at $\rho_0^{exp}=0.277\ \sigma^{-3}$ in
different approximations.}
\begin{tabular}{lccc}
              & HNC/S   &   FHNC/S   &  FHNC/ST   \\  \hline
 $(m_4^{*}/m_4)^{[2]}$             & 1.209 & 1.221 & 1.225  \\
 $(m_4^{*}/m_4)^{[KSA]}$           & 1.197 & 1.208 & 1.213  \\
 $(m_4^{*}/m_4)^{[CA]}$            & 1.187 & 1.196 & 1.200  \\
 $(m_4^{*}/m_4)^{[KSA+ABE]}$       & 1.197 & 1.206 & 1.210  \\
\end{tabular}
\end{table}

\begin{table}[]
\caption{$^4$He effective mass as a function of density in FHNC/ST
approximation.}
\begin{tabular}{lcccc}
 $\rho(\sigma^{-3})$      & 0.253   &  0.277   &  0.300  & 0.330
 \\    \hline
 $(m_4^{*}/m_4)^{[2]}$             & 1.20 & 1.22 & 1.25 & 1.28 \\
 $(m_4^{*}/m_4)^{[KSA+ABE]}$       & 1.19 & 1.21 & 1.23 & 1.26  \\
\end{tabular}
\end{table}


\begin{references}
\bibitem{ED92} D.O. Edwards, M.S. Petersen, and T.G. Culman, J. Low
Temp. Phys. {\bf 89}, 831 (1992).

\bibitem{NA90} M. Nakamura, G. Shirota, T. Shigematsu, K. Nagao, Y.
Fujii, M. Yamaguchi, and T. Shigi, Physica B {\bf 165 \& 166}, 517
(1990).

\bibitem{Saam} W.F. Saam and J.P. Laheurte, Phys. Rev. A {\bf 4}, 1170
(1971).

\bibitem{LA73} J.P. Laheurte, J. Low Temp. Phys. {\bf 12}, 127 (1973).


\bibitem{BF94} J. Boronat, F. Arias de Saavedra, E. Buend\'\i a, and A.
Polls, J. Low Temp. Phys. {\bf 94}, 325 (1994).

\bibitem{FC56} R.P. Feynman and M. Cohen, Phys. Rev. {\bf 102}, 1189
(1956).

\bibitem{OW81} J.C. Owen, Phys. Rev. B {\bf 23}, 5815 (1981).

\bibitem{FF86} A. Fabrocini, S. Fantoni, S. Rosati, and A. Polls,
Phys. Rev. B {\bf 33}, 6057 (1986).

\bibitem{MF83} E. Manousakis, S. Fantoni, V.R. Pandharipande, and Q.N.
Usmani, Phys. Rev. B {\bf 28}, 3770 (1983).

\bibitem{VB88} M. Viviani, E. Buend\'\i a, S. Fantoni, and S. Rosati,
Phys. Rev. B {\bf 38}, 4523 (1988).

\bibitem{Kal} K.E. Schmidt, M.A. Lee, M.H. Kalos, and G.V. Chester,
Phys. Rev. Lett. {\bf 47}, 807 (1981).

\bibitem{AZ79} R.A. Aziz, V.P.S. Nain, J.S. Carley, W.L. Taylor, and G.
T. McConville, J. Chem. Phys. {\bf 70}, 4330 (1979).

\bibitem{ARIAS} F. Arias de Saavedra and E. Buend\'\i a, Phys. Rev.
B {\bf 46}, 13934 (1992).

\bibitem{Leg} A.J. Leggett, Ann. Phys. {\bf 46}, 76 (1968).

\bibitem{Expm3} R.A. Sherlock and D.O. Edwards, Phys. Rev. A {\bf 8},
2744 (1973).

\bibitem{BPF92} J. Boronat, A. Polls and A. Fabrocini,
J. Low Temp. Phys. {\bf 91}, 275 (1993).

\bibitem{Gre} D. Greywall, Phys. Rev. B {\bf 27}, 2747 (1983).







\end{references}
\end{document}